\documentclass[showpacs, oneside, twocolumn, prd, amsmath, amssymb, nofootinbib, superscriptaddress]{revtex4-2}
\usepackage{cases}
\usepackage{amsmath}
\usepackage{amssymb}
\usepackage{amsfonts}
\usepackage{amssymb}
\usepackage{dcolumn}
\usepackage{bm}
\usepackage{bbm}
\usepackage{graphicx}
\usepackage{xcolor}
\usepackage{array}
\usepackage{subfigure}
\usepackage{wasysym}
\usepackage{mathtools}
\usepackage{graphicx}
\usepackage{hyperref}
\hypersetup{colorlinks=true,
	breaklinks=true,
	pdfstartview=Fit,
	linkcolor=myblue,
	citecolor=myred,
	urlcolor=blue
}
\usepackage{cleveref}

\definecolor{myblue}{RGB}{0, 100, 200}
\definecolor{myred}{RGB}{214, 39, 40}
\definecolor{mybrown}{RGB}{123, 64, 26}
\definecolor{mydarkblue}{RGB}{44, 77, 118}

\begin{document}



\title{Modulating binary dynamics via the termination of black hole
    superradiance}

\author{Kaiyuan Fan}
\email{kfanae@connect.ust.hk}
\affiliation{Department of Physics, The Hong Kong University of Science and Technology,\\Clear Water Bay, Kowloon, Hong Kong, P.R. China}
\affiliation{The HKUST Jockey Club Institute for Advanced Study, The Hong Kong University of Science and Technology, Clear Water Bay, Kowloon, Hong Kong, P.R. China}

\author{Xi Tong}
\email{xt246@cam.ac.uk}
\affiliation{Department of Applied Mathematics and Theoretical Physics, University of Cambridge,\\Wilberforce Road, Cambridge, CB3 0WA, UK}
\affiliation{Department of Physics, The Hong Kong University of Science and Technology,\\Clear Water Bay, Kowloon, Hong Kong, P.R. China}
\affiliation{The HKUST Jockey Club Institute for Advanced Study, The Hong Kong University of Science and Technology, Clear Water Bay, Kowloon, Hong Kong, P.R. China}

\author{Yi Wang}
\email{phyw@ust.hk}
\affiliation{Department of Physics, The Hong Kong University of Science and Technology,\\Clear Water Bay, Kowloon, Hong Kong, P.R. China}
\affiliation{The HKUST Jockey Club Institute for Advanced Study, The Hong Kong University of Science and Technology, Clear Water Bay, Kowloon, Hong Kong, P.R. China}

\author{Hui-Yu Zhu}
\email{hzhuav@connect.ust.hk}
\affiliation{Department of Physics, The Hong Kong University of Science and Technology,\\Clear Water Bay, Kowloon, Hong Kong, P.R. China}
\affiliation{The HKUST Jockey Club Institute for Advanced Study, The Hong Kong University of Science and Technology, Clear Water Bay, Kowloon, Hong Kong, P.R. China}

\begin{abstract}
	A superradiant cloud of ultralight bosons near a rotating black hole provides a smoking gun for particle physics in the infrared. However, tidal perturbations from a nearby binary companion can destabilise the boson cloud and even terminate superradiance. In this work, we
    consider the backreaction of superradiance termination to the dynamics of general binary orbits parametrised by their semi-latus rectum, eccentricity and inclination angle. Our analysis focuses on Extreme Mass Ratio Inspiral (EMRI) systems and employs the period-average approximation to derive evolution equations of these binary parameters in the Newtonian limit. We find that the binary evolution history can be significantly modulated by the backreaction towards large circular equatorial orbits with reduced termination rate. This process can generically happen even away from the resonance bands. Our work therefore serves as a first step towards probing ultralight bosons through the statistics of EMRI binary parameters in the future.

\end{abstract}


\maketitle


\section{Introduction}
A rotating Kerr Black Hole (BH) can amplify bosonic fields by giving off its own angular momentum, a well-known mechanism called BH superradiance instability \cite{zel1971generation,brito2020superradiance,zel1972amplification,arvanitaki2010string,Damour:1976kh,Press:1972zz,Cannizzaro:2023jle}\footnote{See \cite{Rahmani_2020,Khodadi_2022,Khodadi_2021,khodadi2020black,Kumar_Jha_2023} for the extensively studied under various conditions, including in higher-dimensional charged-AdS BHs, rotating regular BHs, and alternative gravity theories like f(R) and Horndeski models.}. If the boson possesses a finite mass $\mu$, once a mode has been amplified by the dissipative ergosphere, it can be reflected by the mass barrier on the radial effective potential, and become amplified again at the ergosphere. Such a cycle of amplification and reflection leads to the exponential growth of the energy density stored in this mode, resulting in the formation of a bosonic cloud around the BH. In the non-relativistic limit, the cloud resembles the electron cloud in a hydrogen atom, thereby giving rise to the name Gravitational Atom (GA) for this BH-cloud system \cite{arvanitaki2010string,arvanitaki2011exploring}. 

The GA can be observed through various channels.  For example, the GA emits monochromatic Gravitational Waves (GWs) with a frequency twice the bosonic mass \cite{arvanitaki2011exploring,yoshino2014gravitational,Chan:2022dkt}. These GWs can in principle be detected by ground-based or space-based GW detector, such as the Laser Interferometer Gravitational-Wave Observatory (LIGO) and the Laser Interferometer Space Antenna (LISA) \cite{Brito:2017zvb,LISAConsortiumWaveformWorkingGroup:2023arg}. On the other hand, the null detection of these GW signals also constrains the mass of the light boson \cite{Palomba:2019vxe,LIGOScientific:2021rnv,LIGOScientific:2021sio,KAGRA:2022osp,Yuan:2022bem}. Since cloud growth drains angular momentum from the BH, the BH will spin down until reaching the amplification threshold $J_c/M^2$, where the cloud growth is saturated mode-by-mode. Since the amplification threshold of BH spin depends on the BH mass $M$ as well as the boson mass $\mu$, measuring the statistical distribution of BHs on the so-called BH Regge plane ($J$-$M^2$) can shed light on the properties of boson and the GA \cite{Arvanitaki:2016qwi,Brito:2017zvb,Ng:2020ruv,Hui:2022sri}. If the ultralight boson is an axion, it can couple to electromagnetic fields and plasmas \cite{Yoo:2021kyv,Spieksma:2023vwl}, while also producing observable electromagnetic signals \cite{Shakeri:2022usk,Chen:2022kzv}. Moreover, for BHs in binary systems, which are typically the case in reality, the GA phenomenology becomes even richer. Due to the periodic tidal perturbation induced by the binary companion, the GA can undergo resonant transitions when the gap between atomic energy levels matches the orbital frequency of the binary motion \cite{Baumann:2018vus,Baumann:2019ztm,Baumann:2021fkf,Berti:2019wnn}. These Gravitational Collider Physics (GCP) transitions then generate backreaction to the binary orbital motion \cite{Baumann:2019ztm,Takahashi:2021yhy,Takahashi:2023flk} \footnote{See also \cite{Brito:2023pyl} for a fully relativistic treatment.}, which can be detected through GW observations and pulsar-timing techniques \cite{Ding:2020bnl,Tong:2021whq}. One intuitive  understanding of the backreaction is that the tidal effect deforms the cloud and creates a bulge that \textit{leads} the motion of the companion, which then channels angular momentum transfer between the cloud and the orbit \cite{Zhang:2018kib}. Besides, when the companion moves through the cloud, dynamic friction can also affect the binary motion \cite{Ostriker:1998fa,Hui:2016ltb,Zhang:2019eid,Cao:2023fyv}. Tidal perturbations can also couple bound states of the GA to scattering states, leading to the ionisation of boson clouds similar to the photoelectric effect in atomic physics \cite{Baumann:2021fkf,Baumann:2022pkl,Tomaselli:2023ysb}. The mass quadrupole moment of the GA can induce orbital precession even at the Newtonian level, which can be detected through GWs probes or pulsar timing \cite{Su:2021dwz}. In addition, if two GAs form a binary, GCP resonance can induce mass transfer between the primary and the secondary \cite{Guo:2023lbv}. When the binary separation is close to the cloud radius, molecular structures can form with interesting beating patterns \cite{Ikeda:2020xvt,Liu:2021llm}.

However, note that all the phenomena discussed above demand the existence of a boson cloud, which may not be a robust assumption in general. In a previous work \cite{Tong:2022bbl}, we have shown that superradiance can be terminated for BHs with a binary companion\footnote{In fact, the cloud may be destabilised not only by a binary companion, but also by inhomogeneities in the accretion disk \cite{Du:2022trq}.} due to its tidal perturbations. The non-resonant mixing between the superradiant mode and the highly absorptive mode can completely terminate superradiance when the binary separation is smaller than the critical distance, meaning that no new boson cloud can form. Meanwhile, any existing cloud modes will eventually decay and become absorbed by the BH. Angular momentum lost in this process is then transferred to the binary companion in the form of backreaction that is observable via various channels.

However, our previous analysis assumes large circular equatorial orbits, which is only a crude approximation of binary systems in reality. As a result, in this work, we set out to analyse general orbits with non-zero eccentricity $e$ and inclination angle $\iota_*$ (see FIG.~\ref{GeneralOrbitGeometry} for illustration). We compute the correction to the superradiance rate within both the static approximation and the average co-rotation approximation. We then solve the evolution of binary parameters taking into account the superradiance termination backreaction effects. We find that similar to the case of a circular orbit, the loss in the cloud energy in the GA is accompanied by the gain in the energy of the binary companion, typically with increased binary separation. The superradiance termination effect is then weakened correspondingly. One can think of the cloud as effectively resisting the termination effect by pushing away the binary companion. Additionally, termination backreaction also tends to decrease the eccentricity and inclination\footnote{In contrast, GW emission only affects the eccentricity but is blind to the inclination angle to the leading order.}, driving the orbit towards large circular equatorial ones. We demonstrate the time evolution of the orbital parameters in EMRI systems, which sheds light on future analysis for discovering ultralight bosons via the final-state statistics of EMRI systems.  

This paper is organised as follows. In Sect.~\ref{foundamental}, we provide an introduction to the theoretical background and formulae pertaining to the system of GA in a binary. In Sect.~\ref{effect}, we calculate the period-averaged effective superradiance rate in different regimes and approximations. In Sect.~\ref{backreaction}, we examine the backreaction caused by the termination effect on the companion and study how it affects the binary separation, eccentricity, and inclination. Finally, in Sect.~\ref{conclu}, we present our concluding remarks. We adopt the $(-,+,+,+)$ metric sign convention and set $G=\hbar=c=1$ throughout the paper. Our conventions and notations largely follow from \cite{Baumann:2019ztm,Tong:2022bbl}.


\section{Gravitational atom in a binary}
\label{foundamental}

In this section, we briefly review the physics of superradiance instability in the presence of binary tidal perturbations. Our treatments closely follow \cite{Baumann:2019ztm,Tong:2022bbl}. 


\subsection{The gravitational atom}\label{GAIntroSect}

Consider a Kerr BH with mass $M$ and spin angular momentum $a$. For clarity, we introduce the dimensionless spin parameter $\tilde{a}\equiv a/M$ with $0\leqslant \tilde{a}\leqslant 1$. The dynamics of an ultralight scalar field\footnote{Although the phenomenon of superradiance instability is universal to all ultralight bosons, we will focus exclusively on spin-0 scalar fields in this work. See also \cite{Pani:2012bp,Fell:2023mtf,Brito:2013wya,Alexander:2022avt,Dias:2023ynv,Jia:2023see} for progress on the superradiance of spinning fields.} is described by a Klein-Gordon equation in Kerr spacetime 
\begin{equation}
	\left(\square_{\rm Kerr}-\mu^2\right)\Phi=0~.
\end{equation}
In the non-relativistic limit, one can factor out the dynamical phase due to the rest mass of the boson and evoke the ansatz
\begin{equation}
	\Phi\equiv\frac{1}{\sqrt{2\mu}}e^{-i\mu t}\psi+\rm c.c ~ ,\label{NRAnsatz}
\end{equation}
where $\mu$ is the rest mass of the boson. In cases where the Schwarzschild radius of the BH is smaller than the Compton wavelength of the boson, $R_S\sim M< \mu^{-1}$, one can treat the bosonic field as a classical wave. Substituting \eqref{NRAnsatz} back into the Klein-Gordon equation and expand in powers of the gravitational fine structure constant $\alpha\equiv M\mu< 1$, one obtains a Schr\"{o}dinger-like equation at leading order,
\begin{equation}
	i\partial_t\psi(t,\mathbf{r})=H_0\psi(t,\mathbf{r}) ~ ,~H_0\equiv -\frac{1}{2\mu}\partial^2_{\mathbf{r}}-\frac{\alpha}{r}+\mathcal{O}(\alpha^2)~, \label{FreeEoM}
\end{equation}
with a Newtonian potential resembling the Coulomb potential of the hydrogen atom. After taking the ingoing boundary condition at the BH outer horizon and the vanishing boundary condition at infinity, one can solve the quasi-bound states $\psi_{nlm}$ as modes labelled by the usual three quantum numbers of the hydrogen atom, i.e. $n,l,m$. To leading order in $\alpha$ (or equivalently at large distances where $r\gg M$), the mode functions read
\begin{equation}
	\psi_{nlm}(r,\theta,\phi)\simeq R_{nl}(r)Y_{lm}(\theta,\phi)e^{-i(\omega_{nlm}-\mu)t} ~ , \label{sol}
\end{equation}
where $R_{nl}$ is given by the hydrogen radial function and $Y_{lm}$ stands for spherical harmonics. The typical boson cloud size for a mode with principal quantum number $n$ is given in units of the Bohr radius $r_1=(\mu\alpha)^{-1}$ by $r_n=n^2 r_1$. One distinction from the solution of the hydrogen atom is that the eigenvalues $\omega_{nlm}$ are now complex, $\omega_{nlm}=E_{nlm}+i\Gamma_{nlm}$, as a consequence of the difference in boundary conditions. The real part $E_{nlm}$ gives the energy level of the mode, which enjoys the form \cite{Baumann:2019ztm,Baumann:2019eav} 
\begin{align}
	\nonumber E_{nlm}&=\mu\bigg[1-\frac{\alpha^2}{2 n^2}-\frac{\alpha^4}{8 n^4}-\frac{(3 n- l-1)\alpha^4}{n^4 (l+1/2)}\\
	&\quad+\frac{2 \tilde{a} m \alpha^5}{n^3 l (l+1/2) (l+1)}+\mathcal{O}(\alpha^6)\bigg] ~ .
\end{align}
The imaginary part of the frequency is due to the BH superradiance instability and can be traced back to the ingoing boundary condition at the horizon. From (\ref{sol}), we see that if $\Gamma_{nlm}>0$, the amplitude of the mode can exponentially grow with time, i.e. $|e^{-i\omega_{nlm} t}|\propto e^{+\Gamma_{nlm} t}$, which is the manifestation of superradiance amplification by a rotating BH. In contrast, if $\Gamma_{nlm}<0$, the corresponding mode will exponentially decay, i.e. $|e^{-i\omega_{nlm} t}|\propto e^{-|\Gamma_{nlm}| t}$, and become absorbed by the BH. Henceforth, we refer to modes with positive imaginary part of frequencies as the superradiant modes and those with negative imaginary part of frequencies as absorptive modes. Using the Detweiler approximation (which is valid up to $\alpha\lesssim0.5$ \cite{Brito2015b}), $\Gamma_{nlm}$ is given by \cite{Detweiler:1980uk}
\begin{align}
	\Gamma_{n00}&=-\frac{4}{n^3}\left(1+\sqrt{1-\tilde{a}^2}\right)\mu\alpha^5~,\\
	\Gamma_{nlm}&=2\tilde{r}_+C_{nl}g_{lm}(\tilde{a},\alpha,\omega)(m\Omega_H-\omega_{nlm})\alpha^{4l+5} ~ ,
\end{align}
with $\tilde{r}_+\equiv1+\sqrt{1-\tilde{a}^2}$ and $\Omega_H\equiv\tilde{a}/[2M(1+\sqrt{1-\tilde{a}^2})]$ is the angular velocity of the outer horizon. The definition of $C_{nl}$ and $g_{lm}$ can be found in \cite{Tong:2022bbl}. For an isolated BH, the superradiant modes will grow and extract angular momentum from the BH until the BH spins down to saturate the superradiance threshold at $m\Omega_H=\omega_{nlm}$, where $\Gamma_{nlm}=0$. Afterwards they slowly deplete via emitting monochromatic GWs at a rate \cite{Brito:2014wla,Yoshino:2013ofa}
\begin{equation}
	\gamma_{nlm}=-B_{nl}\frac{S_{\text{c}}/m}{ M^2}\,\mu\alpha^{4l+10} ~ ,\label{GWdepletionRateFormula}
\end{equation}
where $S_{\text{c}}$ is the cloud angular momentum and the numerical coefficients $B_{nl}$ can be found in \cite{yoshino2014gravitational}. We stress that for an isolated GA, the axial symmetry of the spacetime background implies that modes with different azimuthal quantum number $m$ do not linearly mix into each other. Henceforth, each superradiant modes grow, saturate and deplete separately as stated above. We will see in the next subsection that this is no longer the case when the axial symmetry is broken by a nearby binary companion.


\subsection{Tidal perturbations from a binary companion}
\begin{figure}[htbp]
	\centering
	\includegraphics[width=1.05\linewidth]{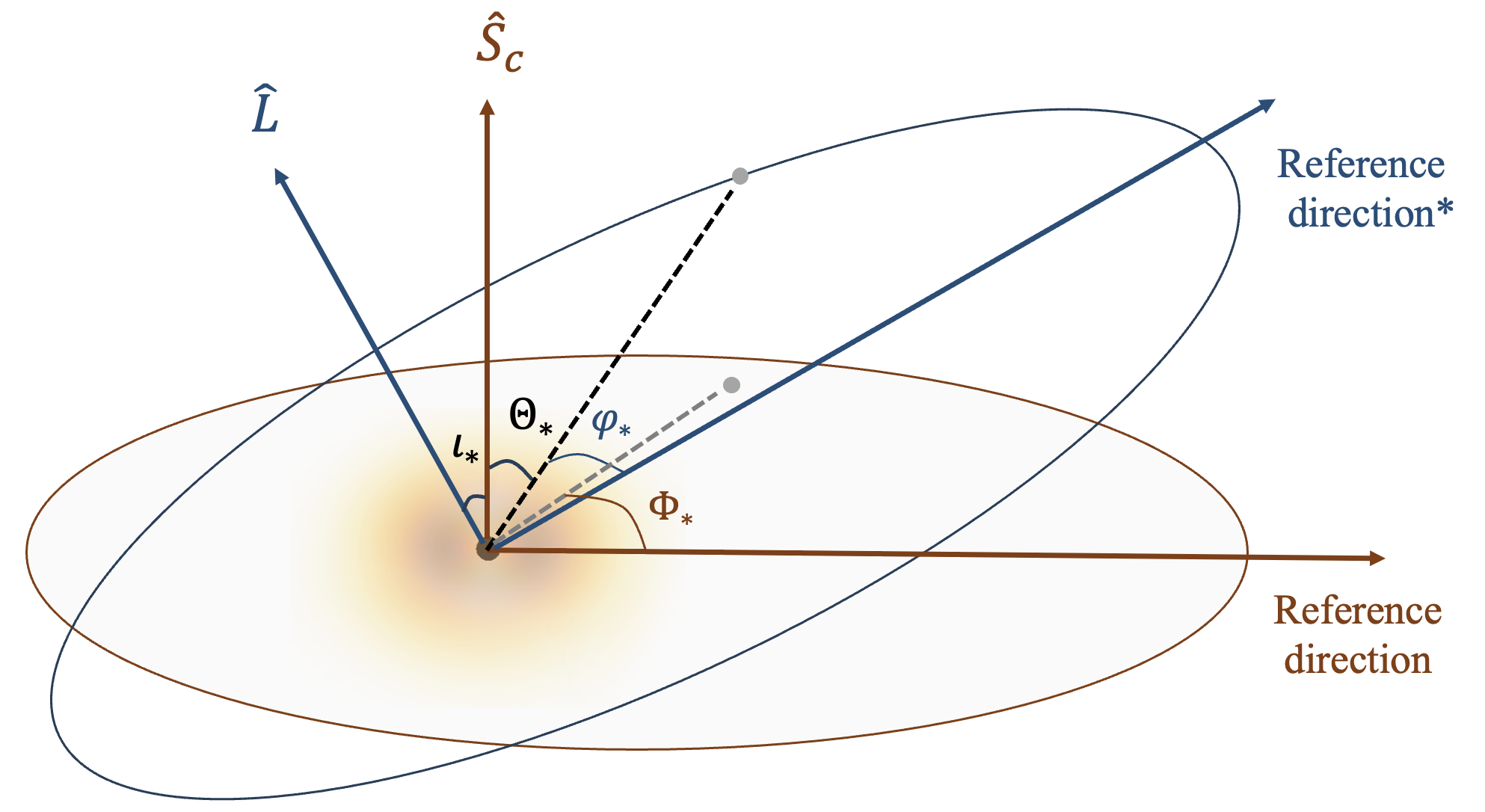}
	\caption{Illustration of Keplerian elements we use to describe the binary system with eccentricity and inclination. To indicate the relative position of the GA and the binary companion, we can introduce two coordinate frames. The orbital frame coordinates (\textcolor{mydarkblue}{blue}) are given by the inclination $\iota_*$, true anomaly $\varphi_*$ and the binary separation $R_*$. The cloud frame coordinates (\textcolor{mybrown}{brown}) are given by the spherical coordinates $(R_*,\Theta_*,\Phi_*)$ defined by the BH spin. Here $\hat{S}_c$ is the spin direction of the GA, and $\hat{L}$ is the direction of angular momentum of the binary companion. Without loss of generality, we choose $0\leqslant\iota_*\leqslant\pi$ and $\dot\varphi_*\geqslant0$ is always positive. Under this choice, $0\leqslant\iota_*\leqslant\pi/2$ gives a co-rotating orbit while $\pi/2\leqslant\iota_*\leqslant \pi$ gives a counter-rotating orbit.}
	\label{GeneralOrbitGeometry}
\end{figure}
Now let us embed the GA into a binary system, where the companion can be either a BH, a neutron star or a white dwarf. We consider a general orbit geometry illustrated in FIG.~\ref{GeneralOrbitGeometry}. The gravity of binary companion induces perturbations to the metric around the boson cloud. In Fermi Normal Coordinates, one can perform a multipole expansion for the Newtonian potential, which starts out at quadrupole order and behaves as a tidal perturbation depending on the mass of the companion $M_*$ and the binary separation $R_*$,
\begin{align}
	\nonumber V_*(r,\theta,\phi)&=-\alpha q\sum_{l_*\geq2}\sum_{|m_*|\leq l_*}\mathcal{E}_{l_*m_*}Y_{l_*m_*}(\theta,\phi)\\
	&\left(\frac{r^{l_*}}{R_*^{l_*+1}}\vartheta(R_*-r)+\frac{R_*^{l_*}}{r_*^{l_*+1}}\vartheta(r-R_*)\right) ~ , \label{VstarDef}
\end{align}
where $\vartheta$ is the Heaviside step-function and $q\equiv M_*/M$ is the mass ratio of the companion and the GA. Here we have also introduced the tidal moments\footnote{Note that there is a typo in Eq. (28) of our previous paper \cite{Tong:2022bbl}, which should be corrected as $\mathcal{E}_{l_*m_*}=4\pi/(2l_*+1)Y^*_{l_*m_*(\pi/2),\varphi_*(t)}=e_{l_*m_*}e^{-im_*\varphi_*(t)}$.}
\begin{equation}
	\mathcal{E}_{l_*m_*}=\frac{4\pi}{2l_*+1}Y^*_{l_*m_*}(\Theta_*,\Phi_*) ~ . 
\end{equation}
The angle $\Theta_*$ represents the angle between the BH spin direction and the companion location, while $\Phi_*$ represents the projection of the true anomaly onto the BH equatorial plane as shown in FIG.~\ref{GeneralOrbitGeometry}. These two angles are related to the inclination angle $\iota_*$ and the true anomaly $\varphi_*$ by
\begin{align}
	\cos\Theta_*&=\sin\iota_*\,\cos\varphi_*~,\\
	\tan\Phi_*&=\sec\iota_*\,\tan\varphi_*~ .
\end{align}
Notice that $\Theta_*(t)$, $\Phi_*(t)$ and $\varphi_*(t)$ are strongly time-dependent during an orbital period of the binary, whereas the inclination angle $\iota_*(t)$ stays nearly constant during any single orbital period, but can pick up a slow time dependence in the long-term evolution (which will be shown later in Sect.~\ref{backreaction}). 

This additional Newtonian potential $V_*$ enters the boson field's equation-of-motion \eqref{FreeEoM} and induces overlaps between any two $H_0$-eigenstates in the form
\begin{align}
	\nonumber\langle\psi_{n'l'm'}|V_*(t)|\psi_{nlm}\rangle&\equiv(-1)^{m'+1}\alpha q\\
	&\sum_{l_*m_*}\mathcal{E}_{l_*m_*}(t)\mathcal{G}^{l'l_*l}_{-m'm_*m}I_r ~ .
	\label{pertur}
\end{align}
Here $I_r$ represents the radial integral
\begin{align}
	\nonumber I_r=&\,\int_0^{R^*} r^2{\rm{d}}rR^*_{n'l'}(r)R_{nl}(r)\frac{r^{l_*}}{R_*^{l_*+1}}\\
	&+\int_{R^*}^\infty r^2{\rm{d}}rR^*_{n'l'}(r)R_{nl}(r)\frac{R_*^{l_*}}{r_*^{l_*+1}} ~ .
\end{align}
We will be mostly working with the scenario where the binary companion is outside the cloud of the GA, thus $I_r$ is dominated by the first term. The Gaunt integral
\begin{equation}
	\mathcal{G}^{l'l_*l}_{-m'm_*m}=\int{\rm{d}}\Omega Y_{l'-m'}(\Omega)Y_{l_*m_*}(\Omega)Y_{lm}(\Omega)
\end{equation}
is non-vanishing only when the following selection rules are satisfied:
\begin{equation}
	\left\{~\begin{aligned}
		&-m'+m_*+m=0~,\label{SelectionRules}\\
		&l+l_*+l'=2k,\text{ for } k\in\mathbb{Z}~,\\
		&|l-l'|\leqslant l_*\leqslant l+l'~.
	\end{aligned}\right.
\end{equation}
When the orbital frequency matches the energy difference between two $H_0$-eigenstates of the cloud, the GA undergos a resonant mode transition guided by the selection rules above. These resonant transitions are sometimes called GCP transitions \cite{Baumann:2019ztm}. These atomic transitions of cloud modes in turn backreact on the binary orbit, which can be detected through multiple observational channels \cite{Baumann:2019ztm,Ding:2020bnl,Tong:2021whq}.

 
\section{Effective superradiance rate for general orbits}
\label{effect}
The mode transitions occur not only in a resonant fashion at specific orbital frequencies, but also in a non-resonant fashion at arbitrary orbital frequencies. In particular, if a superradiant mode transits to an absorptive mode due to the overlap caused by the tidal perturbation, the effective growth rate of the superradiant mode will be decreased. For a superradiant mode $|\psi_{nlm}\rangle$, tidal perturbations generically couple it to the spherically symmetric mode\footnote{One exception is the $|\psi_{n11}\rangle$ modes, where the selection rules \eqref{SelectionRules} forbid their overlap with $|\psi_{n00}\rangle$.} $|\psi_{n00}\rangle$, which possess a large absorption rate. Thus even if the tidal perturbations are small for a large binary separation, the correction from mode mixing can still be considerable and may even overturn the original growth rate of $|\psi_{nlm}\rangle$, terminating superradiance. This superradiance termination effect and its backreaction to the orbital motion of the binary have been previously discussed in \cite{Tong:2022bbl}, but was restricted to large circular equatorial orbits. In this section, we will generalise the discussion to more general orbits with non-zero eccentricity and inclination angle.


\subsection{The adiabatic case and the static approximation}
We start with the perturbed Hamiltonian $H=H_0+V_*$ consisting of a free part $H_0$ (see \eqref{FreeEoM}) and a tidal perturbation part $V_*$ (see \eqref{VstarDef}),
\begin{equation}
	\langle \psi_{n'l'm'}|H|\psi_{nlm}\rangle=\omega_{nlm}\delta_{n'n}\delta_{l'l}\delta_{m'm}+\langle \psi_{n'l'm'}|V_*|\psi_{nlm}\rangle ~.\label{HamiltonianGeneralForm} 
\end{equation}
The diagonal terms are led by the $H_0$ eigenvalues $\omega_{nlm}$, while the off-diagonal terms represent mode mixings introduced by $V_*$. For simplicity, we restrict ourselves to a two-mode subspace $\{|1\rangle,|2\rangle\}$ with a superradiant mode denoted as $|1\rangle\equiv|\psi_{nlm}\rangle$ and an absorptive mode denoted as $|2\rangle\equiv|\psi_{n'l'm'}\rangle$. Defining the mixing coupling as $\eta\equiv V_{21}\equiv\langle2|V_*|1\rangle$, we can express the Hamiltonian \eqref{HamiltonianGeneralForm} in matrix form as 
\begin{align}
	H=   \left(
	\begin{matrix}
		\omega_1+V_{11} & V_{12} \\
		V_{21} & \omega_2+V_{22}
	\end{matrix}
	\right)
	=
	\left(
	\begin{matrix}
		\bar{E}_1(t)+i\Gamma_1 & \eta^*(t) \\
		\eta(t) & \bar{E}_2(t)+i\Gamma_2
	\end{matrix}
	\right) ~ ,
	\label{perturbedHamiltonian}
\end{align} 
where $\bar{E}_i(t)\equiv E_i+V_{ii}(t)$, $i=1,2$, and we have explicitly spelled out the time dependence. Throughout this work, we assume that the companion is light and far away from the GA, i.e. $q\ll 1$ and $\min_t R_*(t)\gg r_n$. Consequently the real part of the diagonal terms are dominated by the free energy levels, i.e. $\bar{E}_i(t)\approx E_i$. The off-diagonal terms, however, cannot be discarded even if they are small, since they play the leading role of mixing $H_0$-eigenstates. More specifically, we assume the hierarchy $|V_{ij}|\ll |E_i-E_j|\ll E_i$. In the adiabatic limit where
\begin{align}
	\left|\frac{\dot{\eta}/\eta}{E_1-E_2}\right|\ll 1~,
\end{align}
one can apply the WKB approximation and solve the occupation coefficient of the cloud $c_i(t)\equiv \langle i|\psi(t)\rangle$ by
\begin{equation}
	c_i(t)=C_{i+}e^{-i\int\lambda_+ \rm{d}t}+C_{i-}e^{-i\int \lambda_- \rm{d}t}~,~i=1,2~,\label{naiveWKB}
\end{equation}
where $\lambda_\pm$ are the instantaneous eigenvalues of the perturbed Hamiltonian,
\begin{align}
	\nonumber \lambda_\pm\equiv
	&\frac{\bar\omega_1+\bar\omega_2}{2}\pm\sqrt{|\eta|^2+\left(\frac{\bar\omega_1-\bar\omega_2}{2}\right)^2}\\
	\simeq&\left\{
		\begin{array}{lr}
			 \bar E_1+\frac{|\eta|^2}{\bar E_1-\bar E_2}+i\left[\Gamma_1-\frac{\Gamma_1-\Gamma_2}{(\bar E_1-\bar E_2)^2}|\eta|^2\right] \text{, } +&\\\\
			 \bar E_2+\frac{|\eta|^2}{\bar E_2-\bar E_1}+i\left[\Gamma_2-\frac{\Gamma_2-\Gamma_1}{(\bar E_1-\bar E_2)^2}|\eta|^2\right] \text{, }- &
		\end{array}
	\right.~.
\end{align}
Here in the second line we have expanded $\lambda_\pm$ in powers of $|\eta/(E_1-E_2)|\ll 1$ and truncated to the first non-trivial order. In summary, the correction to the superradiance rate is given by
\begin{equation}
	\Delta\Gamma_1(t)\simeq-\frac{\Gamma_1-\Gamma_2}{(E_1-E_2)^2}\left|\eta(t)\right|^2 ~ .\label{correction}
\end{equation}
The validity of \eqref{correction} rests upon the adiabatic limit and the WKB approximation. Physically it means when the binary motion is slow, one can neglect the motion of the companion, and obtain the effective superradiance rate as dictated by the binary configuration at the given moment in time. Then one can adiabatically vary the binary configuration over an orbital period to obtain the period-averaged effective superradiance rate as a static approximation,
\begin{equation}
	\overline{\Gamma}_1^{(\text{S})}\equiv \Gamma_1+\overline{\Delta \Gamma}_1^{(\text{S})},~\overline{\Delta\Gamma}_1^{(\text{S})}=\frac1T\int_0^T \Delta\Gamma_1(t) \rm{d}t~.
\end{equation}
To make manifest the time dependence of $\Gamma_1(t)$ and complete the above period-average integral, we introduce a parametrisation of the elliptic orbit as\footnote{In this work, we shall ignore precession effects since they do not enter the dynamics at the level of our approximations. Thus we set the periapsis at $\varphi_*=0$.}
\begin{equation}
	R_*(t)=\frac{p}{1+e\cos\varphi_*(t)}~,
\end{equation}
where $e$ is the eccentricity and $p$ is the semi-latus rectum which is related to the semi-major axis $a$ by $p\equiv a(1-e^2)$. Using the total orbital angular momentum $L$ and the total orbital energy $E$ of the binary,
\begin{equation}
	L=\sqrt{M M_* m_r p},~E=\frac{M M_*}{2p}(e^2-1) ~ ,  
\end{equation}
where $m_r\equiv M q/(1+q)$ is the reduced mass, we can write the angular velocity as a function of the true anomaly,
\begin{equation}
	\dot{\varphi}_*=\frac{L}{m_r R_*^2}=\sqrt{\frac{M(1+q)}{p^3}}(1+e\cos\varphi_*)^2 ~ .  
\end{equation}
In this way we can rewrite the period-average integral as an average over the angle $\varphi_*$,
\begin{equation}
	\overline{\Delta\Gamma}_1^{(\text{S})}=\frac1T\int^{2\pi}_0\Delta\Gamma_{1}\frac{{\rm{d}}\varphi_*}{\dot{\varphi}_*} ~ , \label{DeltaGammaStatic}
\end{equation}
where $T=2\pi\sqrt{p^3/[M(1+q)(1-e^2)^3]}$ is the orbital period of the binary. For the $|1\rangle=|\psi_{322}\rangle$ mode, to leading order in $\alpha$, we found
\begin{align}
	\nonumber\overline{\Delta\Gamma}_{322}^{(\text{S})}&=-\frac{q^2}{\alpha^{10}}\frac{M^5}{p^6}\left(1-\sqrt{1-\tilde{a}^2}\right)\left(1-e^2\right)^{3/2}\\
	\nonumber&\qquad\left[A(e)+B(e)\cos(2\iota_*)+C(e)\cos(4\iota_*)\right]\\
	&\quad\times\left[1+\mathcal{O}(\tilde{a}\alpha,\alpha^{2})\right]~,\label{DeltaGammaAverageStatic}
\end{align}
where
\begin{align}
	A(e)&\equiv \frac{455625}{65536}\left(656+1488 e^2+169 e^4\right)~,\\
	B(e)&\equiv \frac{455625}{16384}\left(80+336 e^2+45 e^4\right)~,\\
	C(e)&\equiv \frac{455625}{65536}\left(48+240 e^2+35 e^4\right)~,
\end{align}
are fourth-order polynomials in the eccentricity. Note that although $|\psi_{322}\rangle$ can mix into absorptive modes other than $|\psi_{300}\rangle$, their contribution to $\overline{\Delta\Gamma}_{322}^{(\text{S})}$ is highly suppressed in $\alpha$. Thus to leading order, \eqref{DeltaGammaAverageStatic} is indistinguishable from the total termination rate. To make a numerical comparison, we note that the superradiant growth rate of $|\psi_{322}\rangle$ at maximal BH spin ($\tilde{a}=1$) and its GW depletion rate at saturation are given by
\begin{align}
	\nonumber\Gamma_{322}&\simeq +\,8.\times10^{2}\,{\rm Myr^{-1}}\left(\frac{\alpha}{0.2}\right)^{13}\left(\frac{M}{10^3M_{\odot}}\right)^{-1}\\&\quad\times\left[1+\mathcal{O}(\alpha)\right]~ , \\
	\nonumber\gamma_{322}&\simeq -\,6.\times10^{-6}\,{\rm Myr^{-1}}\left(\frac{\alpha}{0.2}\right)^{20}\left(\frac{M}{10^3M_{\odot}}\right)^{-1}\\&\quad\times\left[1+\mathcal{O}(\alpha)\right]~.
\end{align}
The superradiance termination rate can be quite significant, too, if the binary separation is small,
\begin{align}
	\nonumber\overline{\Delta\Gamma}_{322}^{(\text{S})}&\simeq -\,5.\times10^{2}\,{\rm Myr^{-1}}\\
	\nonumber&\qquad\left(\frac{\alpha}{0.2}\right)^{-10}\left(\frac{q}{1.4\times 10^{-3}}\right)^2 \left(\frac{M}{10^3M_{\odot}}\right)^{-1}\\&\quad\times\left[1+\mathcal{O}(\alpha)\right]~,
\end{align}
where we have chosen $p=10^3M$, $e=0.5$ and $\iota_*=\pi/4$ as a benchmark.

\subsection{The diabatic case and the average-co-rotation approximation}
\begin{figure}[htbp]
	\centering
	\includegraphics[width=0.85\linewidth]{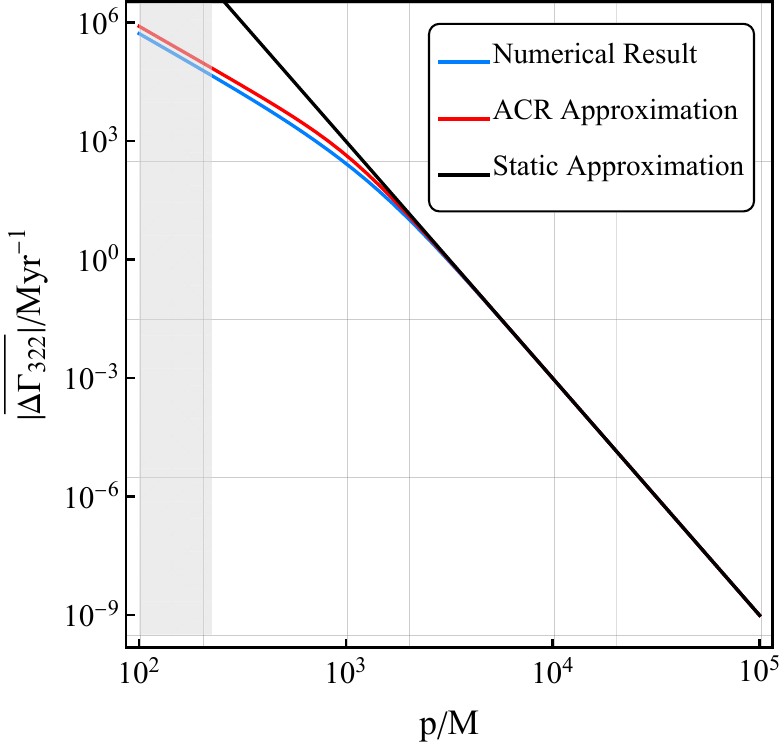}
	\caption{We compare the correction to the superradiance rate due to mode mixing with three different levels of approximations to the superradiance rate. The \textcolor{myblue}{blue} curve shows the numerically computed period-averaged result \eqref{DeltaGammabeforeACR}, the \textcolor{red}{red} curve shows the ACR approximation result \eqref{DeltaGammaafterACR}, and the \textcolor{black}{black} curve shows the static approximation result \eqref{DeltaGammaStatic}. The grey region indicates that the orbit lies inside the boson cloud and must be excluded as it is beyond the validity range of our multipole expansion. Here we chose the superradiant mode $|1\rangle=|\psi_{322}\rangle$ and considered its mixing into the absorptive mode $|2\rangle=|\psi_{300}\rangle$. The parameters are chosen to be $\alpha=0.2$, $M=10^3 M_\odot$, and $q=1.4\times10^{-3}$. The orbit has an eccentricity of $e=0.5$ and an inclination angle of $\iota_*=\pi/4$. Moreover, we have selected the BH spin at the saturation value of $|\psi_{322}\rangle$, i.e. $\tilde{a}_{\text{c}}=2\alpha/(1+\alpha^2)$. Clearly at a large binary separation $p$, the three results all match well. However, at small binary separations where the tidal perturbations are diabatic, there is a significant overestimation for the static approximation whereas the ACR approximation tracks the numerical result very well. The dependence of the superradiance termination rate roughly follows a broken power-law function, which scales as $p^{-6}$ in the adiabatic limit and as $p^{-3}$ in the diabatic limit.}
	\label{figDeltaGammaComparison}
\end{figure}

The WKB approximation we used in the previous discussion requires the adiabaticity of the system. However, if the binary separation is small, the rising orbital frequency increases the time dependence in the mixing coupling, $T^{-1}\sim \left|\dot \eta/\eta\right|$. This leads to the breakdown of adiabaticity,
\begin{align}
	\left|\frac{\dot{\eta}/\eta}{E_1-E_2}\right|\sim \frac{1}{|E_1-E_2|T}\gtrsim 1~.
\end{align}
In such diabatic cases, we need to apply an alternative treatment that takes into account the highly oscillatory mixing coupling. In the circular equatorial orbit case, one can perform a unitary transformation that brings the system into a co-rotating frame in which the time dependence is much weaker and WKB can be safely used \cite{Baumann:2019ztm,Ikeda:2020xvt,takagi1991quantum}. However, for general orbits, it is much less straightforward to find a unitary transformation that accomplishes this goal. One may need a systematic Floquet analysis to fully solve the problem \cite{Baumann:2019ztm}. Therefore, in favour of practicality and physical intuition, we set out on a different route and make a crude approximation by averaging over the instantaneous co-rotating frames. More precisely speaking, for each instant in a binary period, we compute the instantaneous angular velocity of the companion with respect to the GA, and then imagine the companion is in a circular equatorial orbit with the same angular velocity (the separation remains still $R_*$).
For this fictitious circular orbit, we perform a unitary transformation to go into its co-rotating frame 
\begin{align}
	\nonumber&H_D=U(t)^\dagger (H(t)-i\partial_t) U(t),\\
	&\text{with  }U(t)\equiv e^{-i\varphi_*(t) L_z}~,
\end{align}
where $L_z$ is the spin-$z$ direction angular momentum operator for the boson field. Including the selection rules (\ref{SelectionRules}) and the Hamiltonian after the transformation reads 
\begin{equation}
	H_D=\left(
	\begin{matrix}
		\bar{E}_1+i\Gamma_1-m_1 \dot\varphi_* & |\eta| \\
		|\eta| & \bar{E}_2+i\Gamma_2-m_2 \dot\varphi_*
	\end{matrix}
	\right) ~ . 
\end{equation}
Going through the same procedure as in the last subsection, we obtain the correction to the superradiance rate 
\begin{equation}
	\Delta \Gamma_1\simeq -\frac{\Gamma_1-\Gamma_2}{\left[E_1-E_2-(m_1-m_2)\dot\varphi_*(t)\right]^2}|\eta(t)|^2 ~ .\label{2stateDeltaGamma}
\end{equation}
In this case, when $\bar{E}_1-\bar{E}_2\sim (m_1-m_2)\dot{\varphi}(R_*)$, the orbital frequency will enter the GCP resonance band and hit the pole in \eqref{2stateDeltaGamma}. Since binaries are typically more likely to stay out of the resonance band, and we care more about the evolution off the resonance, we perform a ``Wick rotation'' to get rid of the poles on the real energy domain,
\begin{equation}
	\Delta \Gamma_1\simeq -\frac{\Gamma_1-\Gamma_2}{\left(E_1- E_2\right)^2+\left[(m_1-m_2)\dot\varphi_*(t)\right]^2}|\eta(t)|^2 ~ . 
\end{equation}
Finally we perform the same average over a period to obtain
\begin{align}
	\overline{\Delta \Gamma_1}=\frac1T\int^{2\pi}_0\Delta\Gamma_{1}\frac{{\rm{d}}\varphi_*}{\dot{\varphi}_*}~.\label{DeltaGammabeforeACR}
\end{align}
Unfortunately this integral is not analytically solvable. To gain better control over the parametric dependences and speed up numerical solutions later in the next section, we perform a further approximation and replace the time-dependent $\dot{\varphi}_*(t)$ in the denominator by its value taken at the semi-latus rectum,
\begin{align}
	\nonumber \Delta \Gamma_1^{(\text{ACR})}&\approx -\frac{\Gamma_1-\Gamma_2}{\left( E_1- E_2\right)^2+\left[(m_1-m_2)\dot\varphi_*(t)|_{R_*=p}\right]^2}|\eta(t)|^2 \\
	&\approx -\frac{\Gamma_1-\Gamma_2}{\left( E_1- E_2\right)^2+(m_1-m_2)^2\frac{M(1+q)}{p^3}}|\eta(t)|^2~,
\end{align}
which can be considered as a typical characterisation of rotational effect of the binary companion. The final Average Co-Rotation (ACR) approximation for the correction of superradiance rate is therefore
\begin{align}
	\overline{\Delta \Gamma_1}^{(\text{ACR})}=\frac1T\int^{2\pi}_0\Delta\Gamma_{1}^{(\text{ACR})}\frac{{\rm{d}}\varphi_*}{\dot{\varphi}_*}~.\label{DeltaGammaafterACR}
\end{align}
To demonstrate the validity of such an approximation, we numerically solve \eqref{DeltaGammabeforeACR} and compare it to the ACR approximation \eqref{DeltaGammaafterACR} and the static approximation \eqref{DeltaGammaStatic} in FIG.~\ref{figDeltaGammaComparison}. It can be seen that in the adiabatic limit where the binary separation is large, all three results agree well with each other. However, in the diabatic limit where the binary separation is small, the ACR approximation tracks the numerical result \eqref{DeltaGammabeforeACR} well, with only an $\mathcal{O}(1)$ mismatch, whereas the static approximation significantly overestimates the superradiance rate correction.

\begin{figure}[htbp]
\centering
\includegraphics[width=0.33\textheight]{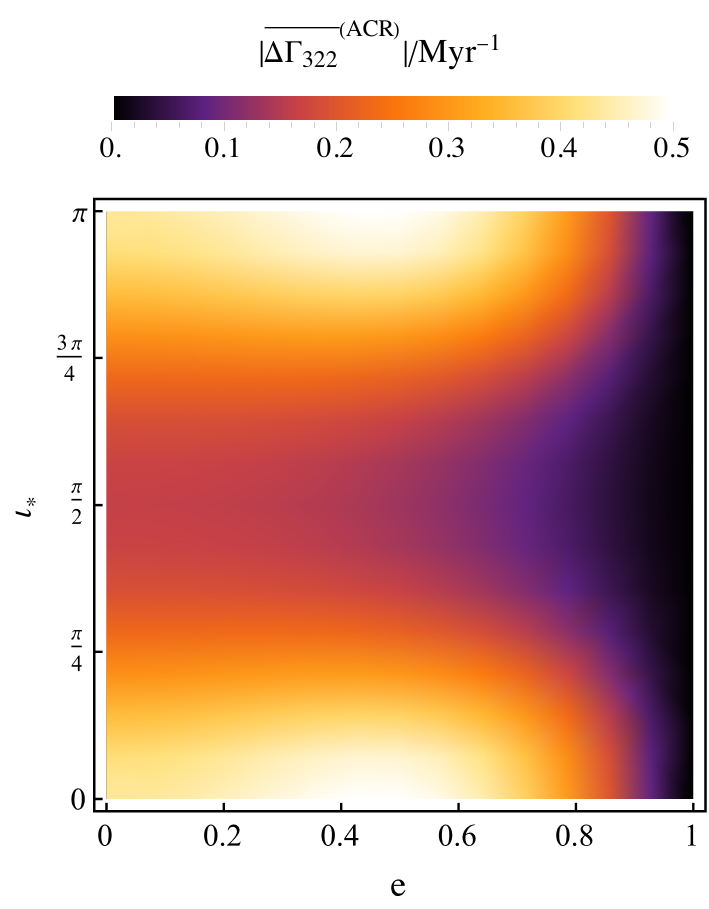}

	\caption{The magnitude of the ACR superradiance termination rate $|\overline{\Delta\Gamma}_{322}^{(\text{ACR})}|$ for the $|\psi_{322}\rangle$ mode as a function of the eccentricity $e$ and inclination angle $\iota_*$. Here we choose $\alpha=0.2$, $M=10^3M_{\odot}$, $q=1.4\times10^{-3}$, $p=4\times10^3 M$, with a BH spin saturated at the $|\psi_{322}\rangle$ threshold.}
	\label{figACREccentricyInclinationDep}
\end{figure}

The ACR superradiance termination rate $\overline{\Delta\Gamma}_1^{(\text{ACR})}(p,e,\iota_*)$ directly depends on the orbital parameters of the binary. The dependence on the semi-latus rectum $p$ is apparently a broken power-law function, as shown in FIG.~\ref{figDeltaGammaComparison}. The dependence on the eccentricity $e$ and the inclination angle $\iota_*$ are more subtle. We plot $\overline{\Delta\Gamma}_{322}^{(\text{ACR})}$ as a function of $e$ and $\iota_*$ with fixed $p$ in FIG.~\ref{figACREccentricyInclinationDep}. We can see with a fixed inclination $\iota_*$, the superradiance termination rate first increases and then drops down to zero with an increasing eccentricity $e$. This is because with the semi-latus rectum $p$ held fixed, increasing $e$ means the decrease of the perigee, which means that the companion is closer to the GA at the periapsis, leading to stronger superradiance termination. However, as the eccentricity continues to increase, the orbit becomes too eccentric with an indefinitely increasing orbital period. Meanwhile, the total amount of suppressed superradiance is bounded from above, suggesting that the period-averaged superradiance termination rate tends to zero as $e$ tends to one. On the other hand, with a fixed eccentricity, the superradiance termination rate is monotonically increasing towards the coplanar configuration with $\iota_*=0,\pi$, and is the smallest when the orbit is orthogonal to the equatorial plane, i.e. $\iota_*=\pi/2$. This is physically understandable because when the companion passes through the spin axis of the GA, the tidal perturbation is axi-symmetric and does not induce any mode mixings (i.e. $\eta=0$). The loss of contribution from this part of the orbit suggests the total effect of superradiance termination is weaker than the coplanar configuration.



\section{Orbital dynamics with termination backreaction}
\label{backreaction}
In the absence of the boson cloud, a binary system gradually loses energy by emitting GWs, which leads to a circularised and shrinking orbit \cite{TGWandEGW}. In the presence of the boson cloud, however, because of the depletion via superradiance termination, the cloud angular momentum is then transferred to the binary companion, thereby also affecting the orbital evolution of the system. In our previous work focused on circular equatorial orbits, such backreaction produces floating or sinking orbits in which the shrinking of binary separation either decelerates or accelerates, for co-rotating orientation and counter-rotating orientation, respectively \cite{Tong:2022bbl}. However, for more general orbits, the consequence of backreaction can be much richer, since all the binary parameters $\{p(t),e(t),\iota_*(t)\}$ can change with time. Physically one can understand the backreaction as the cloud being resistant to the termination effect, exerting an effective torque on the companion that lies along the spin direction of the GA. We will adopt the non-relativistic approximation to solve the evolution of binary parameters. The reason is twofold. First, we focus on the case where the binary separation is much greater than the horizon size of the BH, hence the orbital speed is much smaller than the speed of the light. Second, the relativistic corrections (e.g. the post-Newtonian corrections) do not entail the long-term evolution of the binary parameters more so than the termination backreaction effect, apart from the braking torque from GW emission which we do include in the computation. The precession effect due to post-Newtonian corrections, for instance, do not play a role in affecting the orbital parameters $\{p(t),e(t),\iota_*(t)\}$. To single out the effect of backreaction, we also assume the superradiant mode is saturated (i.e. $\Gamma_1=0$) so that the BH spin does not enter the dynamics.

We start from deriving the effective torque due to the backreaction of the cloud. The total angular momentum of the cloud is
\begin{equation}
	S_{\text{c}}(t)=S_{\text{c},0}\left(m_1|c_1(t)|^2+\sum_{i\neq 1} m_i |c_i(t)|^2\right)~,
\end{equation}
where $m_1 S_{\text{c},0}$ is the saturation cloud angular momentum of superradiance mode $|1\rangle$. For instance, $S_{\text{c},0}\simeq \alpha M^2$ for the mode $|1\rangle=|\psi_{322}\rangle$. Under the saturation assumption, the occupation numbers of other modes are negligible. As a result, we write the cloud angular momentum as $S_{c}(t)= m_1 S_{\text{c},0}|c_1(t)|^2$, which is depleted via two separate channels,
\begin{equation}
	\frac{{\rm{d}}S_{\text{c}}(t)}{{\rm{d}}t}=\left(\frac{{\rm{d}}S_{\text{c}}(t)}{{\rm{d}}t}\right)_{\text{ST}}+\left(\frac{{\rm{d}}S_{\text{c}}(t)}{{\rm{d}}t}\right)_{\text{cGW}} ~ . \label{cloudTotalSpinEq}
\end{equation}
The depletion rate contributed by superradiance termination is 
\begin{equation}
	\left(\frac{{\rm{d}}S_{\text{c}}(t)}{{\rm{d}}t}\right)_{\text{ST}}=2\overline{\Delta\Gamma}_1^{(\text{ACR})}S_{\text{c}}(t) ~ , \label{cloudSTSpinEq}
\end{equation}
and the depletion rate arising from cloud GW emission is given by
\begin{equation}
	\left(\frac{{\rm{d}}S_{\text{c}}(t)}{{\rm{d}}t}\right)_{\text{cGW}}=\gamma_1(S_{\text{c}}) S_{\text{c}}(t) ~ , \label{cloudGWSpinEq}
\end{equation}
where $\gamma_1(S_{\text{c}})$ is given in \eqref{GWdepletionRateFormula}. Notice that the angular momentum carried away by cloud-emitted GWs does not enter binary dynamics at leading order\footnote{The effect of GA mass loss due to GW emission of the cloud is suppressed by the ratio of the cloud mass and the BH mass, which is at least $\mathcal{O}(\alpha^2)$ for the $|\psi_{322}\rangle$ mode. Therefore we can safely neglect this effect at leading order.}, therefore only the depletion due to superradiance termination can give rise to an effective torque backreacting on the binary,
\begin{align}
	\tau_{\text{c}}=-\left(\frac{{\rm{d}}S_{\text{c}}(t)}{{\rm{d}}t}\right)_{\text{ST}}~.
\end{align}
The power of such a torque averaged over a period is
\begin{align}
	P_{\text{c}}=\frac{1}{T}\int {\rm d}t \dot{\varphi}_* \tau_{\text{c}} \cos\iota_*\approx -\frac{2\pi}{T}\left(\frac{{\rm{d}}S_{\text{c}}(t)}{{\rm{d}}t}\right)_{\text{ST}} \cos\iota_*~.
\end{align}
One can understand $P_{\text{c}}$ as the rate of energy injection into the orbital motion of the binary.
Apart from the effect of superradiance termination, the binary system also loses energy and angular momentum through the emission of GWs at a power \cite{TGWandEGW} 
\begin{small}
	\begin{align}
		P_{\rm bGW}=&-\frac{32}{5}\frac{M^5 q^2 (1+q)^{1/2}}{p^5}(1-e^2)\Big(1+\frac{73}{24}e^2+\frac{37}{96}e^4\Big),\label{bGWPower}
	\end{align}
\end{small}
along with a braking torque acting on the binary,
\begin{align}
	\tau_{\rm bGW}=-&\frac{32}{5}\frac{M^{9/2}q^2(1+q)^{1/2}}{p^{7/2}}(1-e)^{3/2}\left(1+\frac78e^2\right)~.\label{bGWTorque}
\end{align}

\begin{figure*}[ht]
		\centering
		\includegraphics[width=1.05\linewidth]{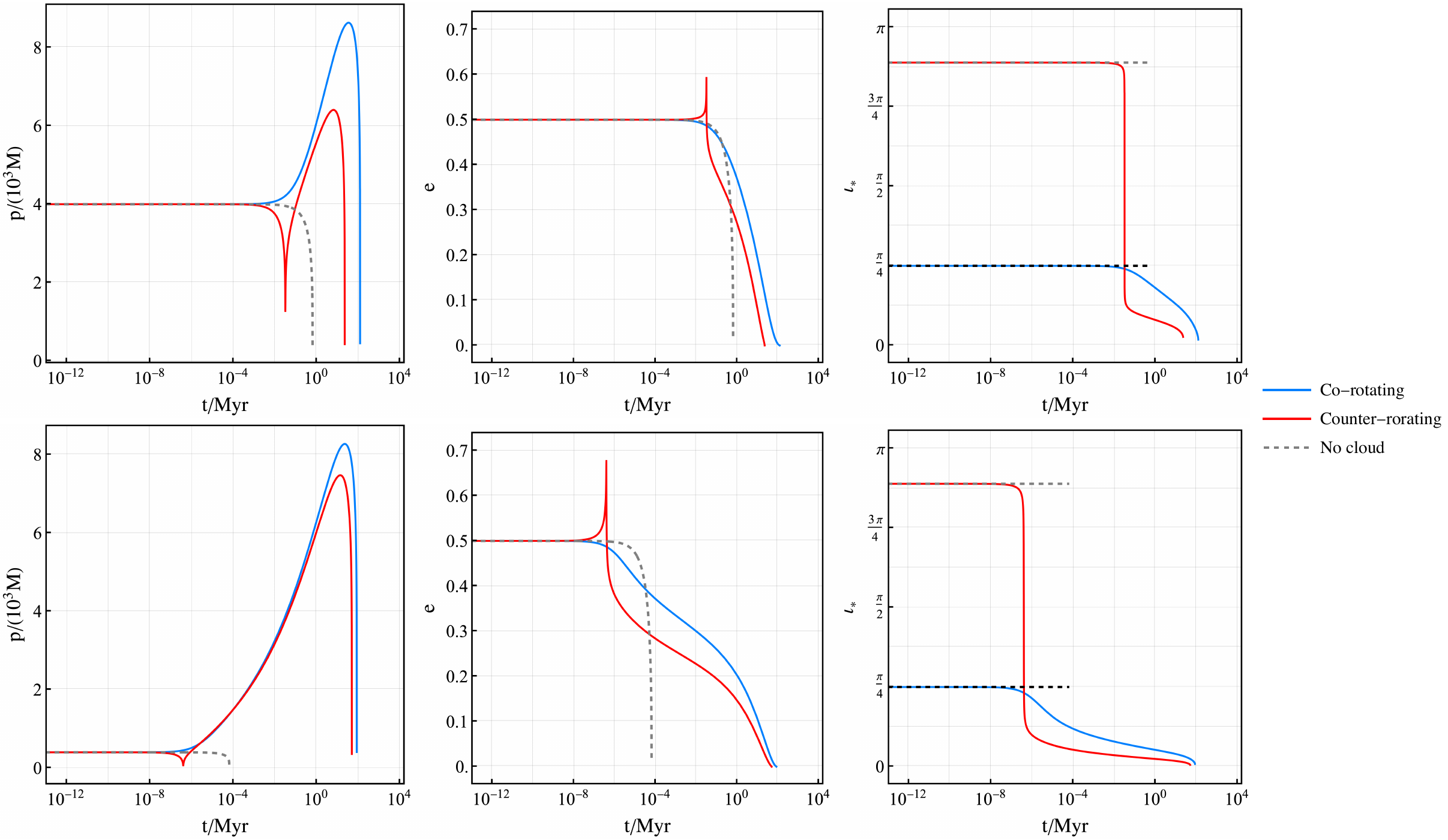}
	\caption{The time evolution of $p(t)$, $e(t)$, and $\iota_*(t)$ in a typical EMRI system with (solid lines) and without (dashed lines) the $|\psi_{322}\rangle$ cloud. \textit{Upper panels}: Large initial binary separation with $p(0)=4\times10^3 M$. \textit{Lower panels}: Small initial binary separation with $p(0)=3\times10^2 M$. The \textcolor{myblue}{blue} lines correspond to co-rotating initial conditions with $\iota_*(0)=\pi/4$, while the \textcolor{red}{red} lines stand for counter-rotating initial conditions with $\iota_*(0)=8\pi/9$. The \textcolor{gray}{grey} dashed lines indicate evolution solely due to GW emission. Other parameters are chosen to be $\alpha=0.2$, $M=10^3 M_\odot$, $q=1.4\times10^{-3}$, and $e(0)=0.5$. Additionally, we assume that the cloud is at the saturation value of the $|\psi_{322}\rangle$ mode, i.e. $\tilde{a}_{\text{c}}=2\alpha/(1+\alpha^2)$. It can be seen from the plots that superradiance termination backreaction tends to bend any binary orbit into a large circular equatorial orbit with the same orientation as the BH spin.}
	\label{figBinaryParameterEvolution}
\end{figure*}

Now we have a system of four variables that evolve in time, namely the three orbital parameters $\{p(t),e(t),\iota_*(t)\})$, together with the cloud spin $S_{\text{c}}(t)$. Thus we need four equations to determine the evolution dynamics. The first equation is given by the cloud spin evolution \eqref{cloudTotalSpinEq}. We also have two scalar equations describing angular momentum conservation projected onto the $\hat{\mathbf{S}}_c$ direction of BH spin as well as its orthogonal plane,
\begin{align}
	\frac{\rm d}{{\rm d} t}[L(t)\,\cos\iota_*(t)]&=\tau_{\text{c}}+\tau_{\rm bGW}\,\cos\iota_*(t)~,\label{angularMomentumEq1}\\
	\frac{\rm d}{{\rm d}t}[L(t)\,\sin\iota_*(t)]&=\tau_{\rm bGW}\,\sin\iota_*(t)~.\label{angularMomentumEq2}
\end{align}
At last, conservation of energy implies
\begin{equation}
	\frac{{\rm d} E(t)}{{\rm d} t}=P_{\text{c}}+P_{\rm bGW}~ . \label{energyConservationEq}
\end{equation}
These four equations, i.e. \eqref{cloudTotalSpinEq}, \eqref{angularMomentumEq1}, \eqref{angularMomentumEq2}, \eqref{energyConservationEq}, when combined with equations relating the orbital energy, angular momenta, power and torques to the four variables $\{p,e,\iota_*,S_{\text{c}}\}$, become a set of first-order ordinary differential equations that completely determines the orbital evolution given the initial conditions,
\begin{align}
	\frac{{\rm d} X_i}{{\rm d} t}=f_i(\{X\})~,\quad X_i\in \{p,e,\iota_*,S_{\text{c}}\}~.\label{collectiveEvolutionEq}
\end{align}
We numerically solve this set of equations and plot the evolution of binary parameters in FIG.~\ref{figBinaryParameterEvolution}.
We give a few remarks on the key features for the evolution of binary parameters and their corresponding physical intuitions below:
\begin{itemize}
	\item From the $p$-$t$ diagram, we observe that superradiance termination backreaction generically pushes the companion further away from the GA, which in turn, reduces the termination rate itself. This is a clear manifestation of the cloud resisting being terminated. For the co-rotating case, $p$ increases monotonically as the cloud pumps energy into the binary, resulting in a floating orbit. For the counter-rotating case, however, $p$ initially decreases, suggesting a short period of sinking orbit, but is eventually pushed away as a floating orbit. The reason for this two-stage evolution will become apparent as we discuss the $\iota_*$-$t$ diagram below. In both cases, at much longer time scales, the cloud eventually depletes and the effect of binary GW emission starts to kick in. The binary separation then drops down, leading to the onset of merger phase. 
	\item From the $e$-$t$ diagram, we can observe that termination backreaction tends to circularise the co-rotating orbit, similar to the effect of binary GW emission. In contrast, the counter-rotating case is again a two-stage evolution in which the eccentricity initially grows but eventually decays.
	\item The $\iota_*$-$t$ diagram shows that termination backreaction universally reduces the inclination angle of the orbit until merger. In other words, the cloud torque tends to align the binary orbit with its spin direction. This solves the mystery of the two-stage evolution in the case with counter-rotating initial condition. Namely, termination backreaction first produces a sinking orbit with increasing eccentricity, then the torque brings down the inclination angle past $\iota_*=\pi/2$, after which the binary companion actually becomes co-rotating with a circularised floating orbit. In summary, termination backreaction tends to bend any binary orbit into a large circular equatorial orbit aligned with the BH spin.
	\item Notice that the grey dashed lines represent the usual binary evolution in the absence of boson cloud. In this case, the orbit is still circularised, but the inclination angle does not change because the equations \eqref{bGWPower} and \eqref{bGWTorque} are blind to BH spin at leading order. Comparison to cases with a cloud shows that the both the merger time and the final-state statistics can be drastically affected by the presence of the cloud. This gives us a potential probe of such boson clouds from the statistics of EMRI systems. We will pursue such an analysis in future works.
\end{itemize}

\begin{figure*}[ht]
	\centering
	\includegraphics[width=17cm]{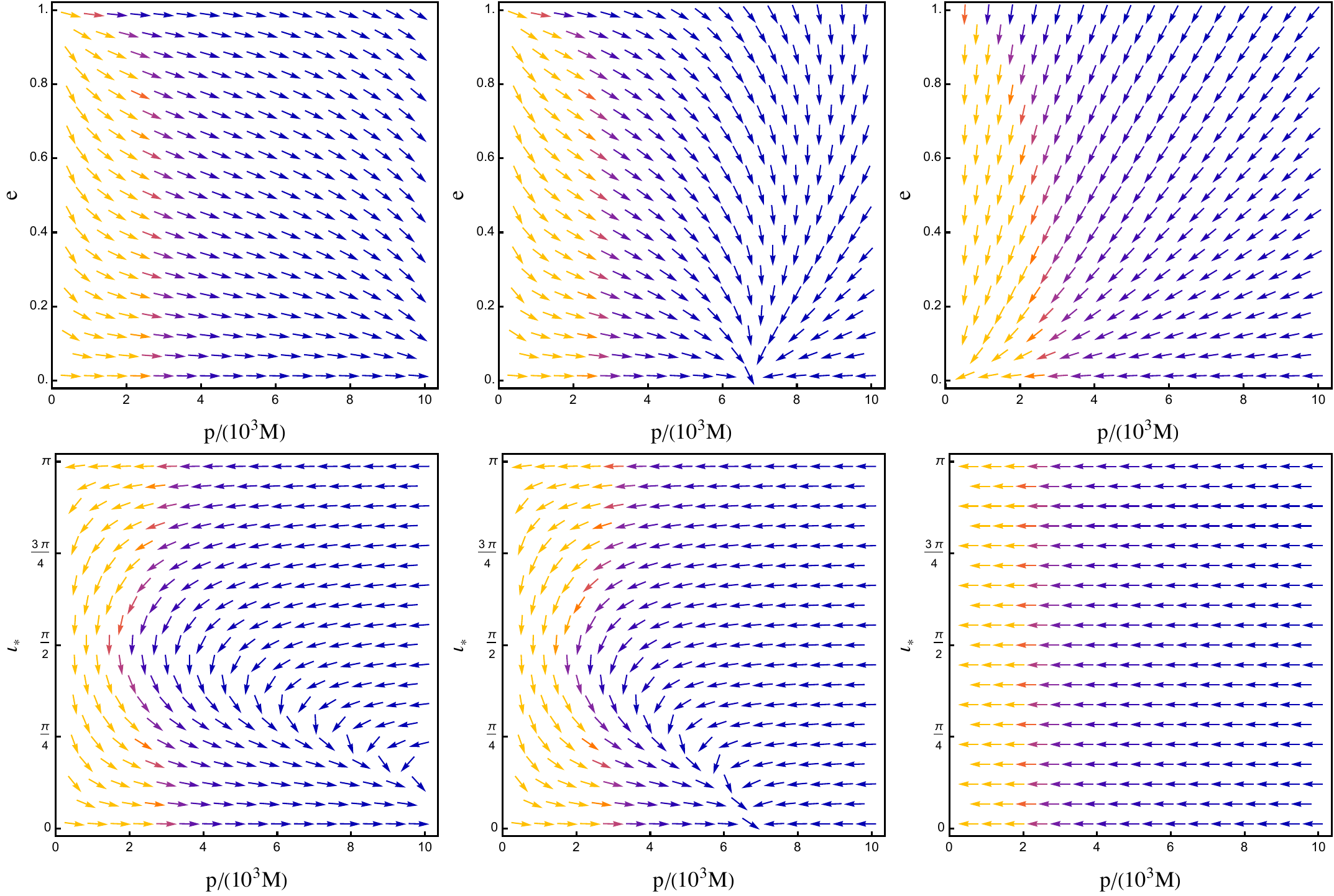}
	\caption{The flow of $p$, $e$, and $\iota_*$ for different sections of constant cloud spin $S_{\text{c}}$ in an EMRI system. Here we choose $\alpha=0.2$, $q=1.4\times10^{-3}$, $M=10^3M_\odot$, together the BH spin set at the threshold of $|\psi_{322}\rangle$. The cloud occupation numbers are chosen to be $|c_{322}|^2=1$ (\textit{left column}), $|c_{322}|^2=0.3$ (\textit{middle column}) and $|c_{322}|^2=0$ (\textit{right column}), which corresponds to cloud spin at $S_{\text{c}}=2S_{\text{c},0}$, $S_{\text{c}}=0.6S_{\text{c},0}$, and $S_{\text{c}}=0$, respectively.}\label{figFlow}
\end{figure*}

One can also analyse the evolution governed by \eqref{collectiveEvolutionEq} as a phase portrait in the parameter space spanned by $\{p,e,\iota_*,S_{\text{c}}\}$. In FIG.~\ref{figFlow}, we plot the gradient vector field $f_i$ which determines the flow of orbital parameters. As it is difficult to visualise a four-dimensional vector field, we take two-dimensional sections by fixing the other two parameters. The first row represents the flow in the $p$-$e$ plane with fixed $\iota_*$ and for three different fixed choices of $S_{\text{c}}$, while the second row depicts the flow in the $p$-$\iota_*$ plane with fixed $e$ and for three different fixed choices of $S_{\text{c}}$. 
A few remarks follow from inspecting the plots:
\begin{itemize}
	\item As expected, cloud backreaction generically decreases eccentricity and inclination while pushing the orbits further away until reaching a transient dynamical balance with effects of binary GW emission.
	\item It is apparent that with a smaller cloud occupation number and thus smaller cloud spin $S_{\text{c}}$, backreaction is weakened correspondingly. In the case with zero cloud occupation, the inclination angle does not evolve at leading order, while the eccentricity still decreases due to binary GW emission.
	\item In some of the plots in FIG.~\ref{figFlow} (e.g. the middle column), there appears to be a fixed-point attractor of time flow. We note that this is an artefact of reducing high-dimensional parameter spaces to two-dimensional sections. Due to energy leakage into outgoing GWs, the binary separation will eventually decrease. The only genuine attractor of this system is the merger event at $p=e=0$.
\end{itemize}

\section{Conclusion}
\label{conclu}
BHs surrounded by an ultralight boson cloud exhibit numerous exciting phenomena, whose discovery can bring invaluable insight into the infrared physics beyond Standard Model. In the meanwhile, however, it is important to test the robustness of postulating such a GA. In cases where the GA is in a binary, superradiance can be terminated by tidal perturbations of the companion. This mechanism has been studied for large circular equatorial orbits in our previous work. In this paper, we have extended our discussions to general orbits with non-trivial eccentricity and inclination. We calculated the effective superradiance rate both in the adiabatic case with the static approximation and in the diabatic case with the average co-rotation approximation. Applying these period-averaged correction to cloud growth rate, we moved on to study the backreaction to the binary orbit in EMRI systems. We found that termination backreaction generically circularises the orbit and pushes it outwards, while also bending the orbital plane towards aligning with the equator of the BH. In particular, even if the companion starts out counter-rotating the GA with a sinking orbit, termination backreaction soon bends the inclination angle into co-rotating and the companion is subsequently pushed outwards as a floating orbit. One can understand this as the cloud being resilient to superradiance termination and tends to give a negative feedback that reduces termination.

There are certainly many directions left to explore in the future. To name a few, although we relaxed the large circular equatorial orbit assumption in this work, many approximations are still evoked to acquire a reasonable analytical formula so as to characterise superradiance termination. It would be more satisfactory to work out a numerical solution of the cloud Schrödinger equation and the equations of motion of the binary, which can offer a more concrete depiction of superradiance termination. It is also interesting to include precession and relativistic corrections and examine how they may affect our predictions. Second, with the control over the evolution history of binary parameters, one can examine the statistical properties of such EMRI systems in general. It is not inconceivable that even if we cannot directly observe the cloud before depletion, its transient presence may be encoded in the final-state statistics of EMRIs. Third, as mentioned in Sect.~\ref{GAIntroSect}, linear mixings between modes with different azimuthal quantum numbers are forbidden by the axial symmetry of the Kerr background. This is not true when one goes to the non-linear regime, where all the modes are expected to couple to each other via local self-interactions or non-local self-gravity \cite{Yoshino:2012kn,Baryakhtar:2020gao,Omiya:2022gwu,Chia:2022udn}. Thus superradiance termination may occur even in the absence of any external perturber, but as a consequence of the interaction of multiple cloud modes that coexist around the BH. We leave this interesting possibility to future works.



\section{Acknowledgment}
HYZ would like to thank Takahiro Tanaka, Hidetoshi Omiya, Takuya Takahashi, and Otto Akseli Hannuksela for the fruitful discussions. This work was supported in part by the National Key R\&D Program of China (2021YFC2203100),  CRF grant C6017-20GF and GRF grant 16306422 by the RGC of Hong Kong SAR. XT is supported by STFC consolidated grants ST/T000694/1 and ST/X000664/1.

\appendix

\bibliography{reference}

\end{document}